\documentclass[preprint,prd,aps,nofootinbib]{revtex4} 
\usepackage{amssymb}
\usepackage{graphicx}
\usepackage{axodraw}

\begin{document}
 
\title {\large Muon anomalous magnetic moment and lepton \\ flavor 
violation in MSSM}
 
\author{ Xiao-Jun Bi }
\email[Email: ]{bixj@mail.tsinghua.edu.cn}
\affiliation{ Department of Physics, Tsinghua University, Beijing 100084,
People's Republic of China}

\author{ Yu-ping Kuang }
\email[Email: ]{ypkuang@mail.tsinghua.edu.cn}
\affiliation{ Department of Physics, Tsinghua University, Beijing 100084,
People's Republic of China}

\author{ Yong-Hong An }
\affiliation{ Department of Physics, Tsinghua University, Beijing 100084,
People's Republic of China}
 
\date{\today}

\begin{abstract}          

We give a thorough analysis of the correlation between the muon anomalous 
magnetic moment and the radiative lepton flavor violating (LFV)
processes within the minimal supersymmetric standard model.
We find that in the case when the slepton mass eigenstates 
are nearly degenerate,
$\delta a_\mu$, coming from SUSY contributions, hardly depends on
the lepton flavor mixing and, thus, there is no direct relation between
$\delta a_\mu$ and the LFV processes. On the contrary, 
if the first two generations'
sleptons are much heavier than the 3rd one, i.e., in the effective
SUSY scenario, the two quantities are closely related.
In the latter scenario, 
the SUSY parameter space to account for the experimental $\delta a_\mu$
is quite different from the case of no lepton flavor mixing.
Especially, the Higgsino mass parameter 
$\mu$ can be either positive or negative.

\end{abstract}
 
\preprint{TUHEP-TH-02139}
 
\maketitle
 
\section {introduction}
Recently, the Brookhaven E821 Collaboration announced their new
experimental result on muon anomalous magnetic moment, $a_\mu=(g_\mu-2)/2$,
with improved statistics\cite{new821},  which is twice precision of 
their 2001 result\cite{old821}.
The present discrepancy between the standard
model (SM) prediction and the measurement, 
depending on the SM hadronic corrections to $a_\mu$, is
\begin{eqnarray}
a_\mu^{exp}-a_\mu^{SM} &=& 26(10)\times 10^{-10}\\
&or&17(11)\times 10^{-10}\ ,
\end{eqnarray}
lying between 1.6$\sigma$ and 2.6$\sigma$. 

Since the first announcement of existing discrepancy between theory
and experiment on $a_\mu$, there appeared a lot of works on this 
subject trying to
explain the result in various extensions of the SM, among which
the most promising new physics is the minimal supersymmetric standard
model (MSSM)\cite{susyg-2}. Although the present E821's measurement 
can not provide
compelling evidence in favor of new physics, it is generally expected
that this deviation will be confirmed when both the experimental and 
theoretical errors are reduced and this result can now be used to put constraint on
the supersymmetric (SUSY) parameters. 

The extensive studies show that 
it is easy to accommodate $\delta a_\mu\sim (10\sim 70)
\times 10^{-10}$ within the MSSM framework if the SUSY particles 
are as `light' as about a few hundred GeV\cite{susyg-2}.
The positive sign of the Higgsino mass parameter, $\mu$,
is strongly favored by the present value of $\delta a_\mu$.
Since in most parameter space the main SUSY contribution to $a_\mu$
comes from exchanging
chargino and scalar muon neutrino virtual particles, which is approximately
proportional
to $\mu M_2\tan\beta$, the sign of $\mu$ is thus positive relative to $M_2$,
the wino mass parameter, provided that SUSY helps to enhance $a_\mu$.

In this work we will study the SUSY contributions to $a_\mu$ in the case
when considering the lepton flavor mixing in the soft breaking sector.
Different from the similar numerical studies of
lepton flavor mixing effects on $a_\mu$\cite{pre},
we will give a thorough analysis of the correlation between
the SUSY contributions to $a_\mu$ and to lepton flavor violation (LFV).
We find in the case that the sleptons and sneutrinos are nearly degenerate,
$\delta a_\mu$ has no direct relation with the LFV processes. Actually,
this is the usual case which has been extensively studied in the
literature\cite{susyg-2}. However, 
it is most important to take into account the effects of lepton flavor
mixing in the case of effective SUSY scenario, where 
$\delta a_\mu$
can only arise when the slepton mixing between the second and the 
third generations is introduced.

We find that in the effective SUSY scenario,
the SUSY parameter space may be quite different from those without
considering the slepton mixing. 
The sign of
$\mu$ can be either positive or negative, to enhance $a_\mu$, 
depending on the lepton flavor
mixing angles. Small $\tan\beta$ is more favored in this case.

The paper is organized as follows.
In the next section we give the analytic expressions for the SUSY
contributions to $a_\mu$ and the branching ratio of LFV processes
$l_i\to l_j\gamma$. In section III, we will present the numerical
results and some approximate upper bound on $\delta a_\mu$.
Finally, we give summary and conclusions in section IV. 

\section{ Analytic expressions }
\subsection{ Expressions of $\delta a_\mu$ and Br$(l_i\to l_j\gamma)$ in MSSM }

\begin{center}
\begin{figure}
\begin{picture}(500,250)(20,50)
\ArrowLine(20,200)(65,200)
\ArrowLine(65,200)(185,200)
\ArrowLine(185,200)(230,200)
\ArrowLine(270,200)(315,200)
\ArrowLine(315,200)(435,200)
\ArrowLine(435,200)(480,200)
\DashArrowArcn(125,200)(60,180,0){4}
\DashArrowArcn(375,200)(60,180,0){4}
\Photon(125,185)(180,130){4}{7}
\Text(150,135)[]{\huge $\gamma$}
\Text(170,250)[l]{\huge $\tilde{\nu}_\alpha$}
\Text(30,185)[]{\huge $\mu(\tau)$}
\Text(220,185)[]{\huge $\mu$}
\Text(125,215)[]{\Large $\chi^-_a$}
\Text(195,125)[]{\huge $q$}

\Photon(380,270)(450,305){4}{7}
\Text(440,285)[]{\huge $\gamma$}
\Text(328,250)[r]{\huge $\tilde{l}_\alpha$}
\Text(280,185)[]{\huge $\mu(\tau)$}
\Text(470,185)[]{\huge $\mu$}
\Text(375,185)[]{\Large $\chi^0_a$}
\end{picture}

\vspace*{-1.5cm}
\caption{\label{fig1} Feynman diagrams of the one-loop SUSY
contribution to $a_\mu$ (and the process $\tau \to \mu\gamma$)
via the exchange of a chargino (left) and via a neutralino (right).}
\end{figure}
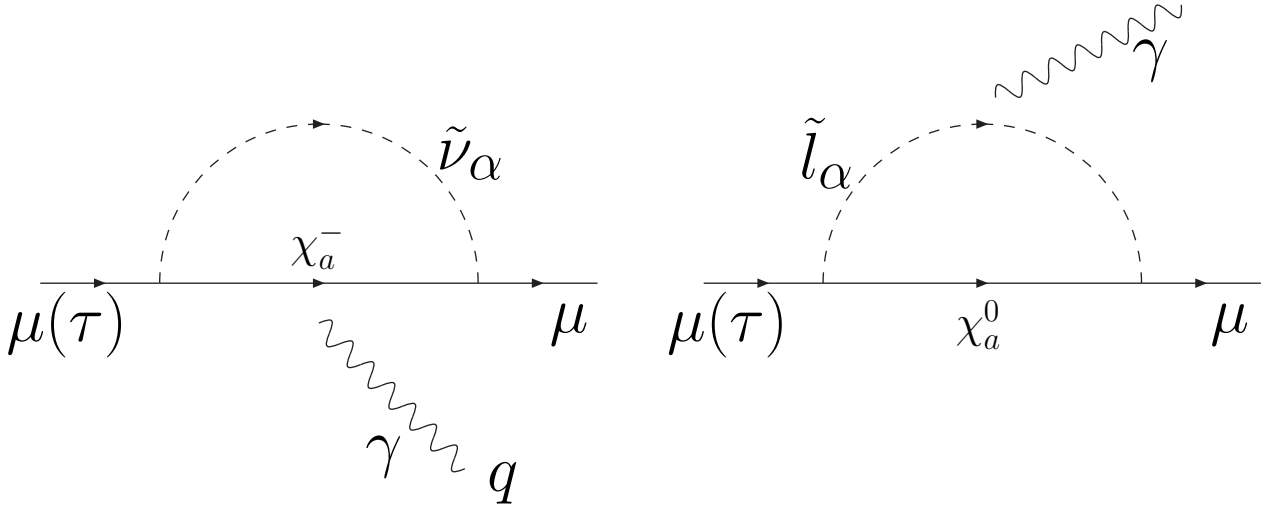
\end{center}

Since the muon's anomalous magnetic moment and the LFV processes 
$l_i\to l_j\gamma$ arise from similar operators,
the effective Lagrangian related to $a_\mu$ and radiative LFV process can
be written in one form,
\begin{equation}
\label{eff}
{\cal L}_{eff}=e\frac{m_i}{2}\bar{l}_j\sigma_{\alpha\beta}F^{\alpha\beta}
(A_L^{ij}P_L+A_R^{ij}P_R)l_i\ ,
\end{equation}
where $P_{L,R}=\frac{1}{2}(1\mp\gamma_5)$ are the chirality projection operators
and $i(j)$ denotes the initial (final) lepton flavor. 
The muon anomalous magnetic moment is given by
\begin{equation}
a_\mu=m_\mu^2(A_L^{22}+A_R^{22})\ ,
\end{equation}
while the branching ratio of 
$l_i\to l_j\gamma$ is given by
\begin{equation}
BR(l_i\to l_j\gamma)=\frac{\alpha_{em}}{4}m_i^5(|A^{ij}_L|^2+|A^{ij}_R|^2)/
\Gamma_i \ \ ,
\end{equation}
with $\Gamma_i$\cite{pdg} being the width of $l_i$.

The SUSY contribution to the form factors $A_L$ and $A_R$ is given by 
the photon-penguin diagrams via exchanging (i) chargino-sneutrino and (ii)
neutralino-slepton, as shown in FIG. \ref{fig1}. 
The analytic expressions for $\delta a_\mu$ from
the neutralino and chargino exchange are
\begin{eqnarray}
\label{auneu}
\delta a_\mu^{(n)}&=&-\frac{1}{32\pi^2}\frac{e^2}{\cos^2\theta_W}
\frac{m_\mu^2}{m_{\tilde{l}_\alpha}^2}\cdot \nonumber\\
&&\left[({A^{i\alpha a}}^{*}A^{i\alpha a}+{B^{i\alpha a}}^{*}B^{i\alpha a})
F_1(k_{\alpha a})+\frac{m_{\chi_a^0}}{m_\mu}
Re({A^{i\alpha a}}^*B^{i\alpha a})F_2(k_{\alpha a})\right]
\\
\text{and}&&\nonumber\\
\label{aucha}
\delta a_\mu^{(c)}&=&\frac{g_2^2}{16\pi^2}
\frac{m_\mu^2}{m_{\tilde{\nu}_\alpha}^2}Z_{\tilde{\nu}}^{i\alpha}
{Z_{\tilde{\nu}}^{i\alpha}}^*\cdot \nonumber\\
&&\left[( {Z_{1a}^+}^*Z_{1a}^++\frac{m_\mu^2}{2M_W^2\cos^2\beta}
{Z_{2a}^-}^*Z_{2a}^- )F_3(k_{\alpha a})
+\frac{m_{\chi_a^-}}{\sqrt{2}M_W\cos\beta}Re(Z_{1a}^+Z_{2a}^-)
F_4(k_{\alpha a})\right] \, , \;\;\;\;\;
\end{eqnarray}
respectively with index $i=2$. In the above expressions the $A$ and $B$ are
the lepton--slepton--neutralino coupling vertices given by
\begin{eqnarray}
\label{ca}
A^{i\alpha
a}&=&\left(Z_{\tilde{L}}^{i\alpha}(Z_N^{1a}+Z_N^{2a}\cot\theta_W)
-\cot\theta_W\frac{m_i}{M_W\cos\beta}Z_{\tilde{L}}^{(i+3)\alpha}
{Z_N^{3a}}\right) \ ,\\
\label{cb}
B^{i\alpha
a}&=&-\left(2Z_{\tilde{L}}^{(i+3)\alpha}{Z_N^{1a}}^*+\cot\theta_W
\frac{m_i}{M_W\cos\beta}Z_{\tilde{L}}^{i\alpha}{Z_N^{3a}}^*\right)
\; ,
\end{eqnarray}
where $Z_{\tilde{L}}$ is the $6\times 6$ slepton mixing matrix and
$Z_N$ is the neutralino mixing matrix. Similarly, $Z_{\tilde{\nu}}$
is the sneutrino mixing matrix, while
$Z^+$ and $Z^-$ are the mixing matrices for the charginos.
The definitions of these mixing matrices and the expressions of
$F_i$'s are given in the appendix.

For the processes $l_i\to l_j\gamma$, the contribution from
 neutralino exchange gives
\begin{eqnarray}
\label{aln}
A_L^{ij(n)}&=&-\frac{1}{32\pi^2}(\frac{e}{\sqrt{2}\cos\theta_W})^2\frac{1}
{m_{\tilde{l}_\alpha}^2}\left[{B^{j\alpha a}}^*B^{i\alpha
a}F_1(k_{\alpha a}) +\frac{m_{\chi_a^0}}{m_i}{B^{j\alpha a}}^*
A^{i\alpha a}F_2(k_{\alpha a})\right]\ ,\\
A_R^{ij(n)}&=&A_L^{(n)}\ (B\leftrightarrow A)\ ,
\end{eqnarray}
while the corresponding contribution coming from chargino exchange is
\begin{eqnarray}
A_L^{ij(c)}&=&\frac{g_2^2}{32\pi^2}{Z_{\tilde{\nu}}^{i\alpha}}^*
Z_{\tilde{\nu}}^{j\alpha}
\frac{1}{m_{\tilde{\nu}_\alpha}^2}\left[ Z^-_{2a}{Z^-_{2a}}^*
\frac{m_im_j}{2M_W^2\cos^2\beta}
F_3(k_{\alpha a})\right. \nonumber \\
&&\left.
+\frac{m_{\chi^-_a}}{\sqrt{2}M_W\cos\beta}Z^+_{1a}Z^-_{2a}
\frac{m_j}{m_i} F_4(k_{\alpha a})\right] \ ,\\
\label{imp}
A_R^{ij(c)}&=&\frac{g_2^2}{32\pi^2}{Z_{\tilde{\nu}}^{i\alpha}}^*
Z_{\tilde{\nu}}^{j\alpha} \frac{1}{m_{\tilde{\nu}_\alpha}^2}\left[
Z^+_{1a}{Z^+_{1a}}^*F_3(k_{\alpha a})
+\frac{m_{\chi^-_a}}{\sqrt{2}M_W\cos\beta}{Z^+_{1a}}^*{Z^-_{2a}}^*
F_4(k_{\alpha a})\right] \; .
\end{eqnarray}

\subsection{ Flavor structure on the interaction basis }

The expressions for $\delta a_\mu$ and Br($l_i\to l_j\gamma$)
in the last subsection show that there is close relations 
between the two quantities.
We notice that all these expressions are given in the
mass eigenstates of the SUSY particles and the lepton
flavor mixing is presented in the mixing matrices on the 
interaction vertices. 
These expressions are suitable for numerical calculations. 
However, to analyze the
flavor structure of the amplitude, 
it is more convenient to work on the {\em interaction
basis}, which is defined as the basis where the lepton mass
matrix and the gauge coupling vertices are all diagonal.
On this basis there are much more Feynman diagrams
than those in FIG. \ref{fig1}. For example, the vertex $A$
given in Eq. (\ref{ca}) actually contains three different 
interaction vertices,
$l_i^L-\tilde{l}_i^L-\tilde{B}$, $l_i^L-\tilde{l}_i^L-\tilde{W}$, 
and $l_i^L-\tilde{l}_i^R-\tilde{H}_D$ on this basis.
Thus only the $A^*A$ term in Eq. (\ref{auneu}) represents
9 different Feynman diagrams.

On this basis the slepton and sneutrino mass matrices are
generally not diagonal. We first give the form of these mass matrices.
The slepton mass matrix can be written in a general form as  
\begin{equation}
\label{mslp}
M_{\tilde{l}}^2=\left(  \begin{array}{cc}
Z_L m_L^2 Z_L^\dagger & -m_l(\mu\tan\beta+ A_l^*) \\
-m_l(\mu^*\tan\beta+ A_l) & 
Z_R m_R^2 Z_R^\dagger  
\end{array} \right)\ \ ,
\end{equation}
where
\begin{eqnarray} 
\label{mslpl}
m_L^2 &=& m_{\tilde{l}}^2 + m_l^2 + \cos2\beta(-\frac{1}{2}+\sin^2\theta_W)M_Z^2\ \ ,\\
\label{mslpr}
m_R^2 &=& m_{\tilde{r}}^2  + m_l^2 - \cos2\beta\sin^2\theta_WM_Z^2\ \ ,
\end{eqnarray} 
and $m_l$ is the diagonal mass matrix of leptons.
Here $m_{\tilde{l}}^2$ and $m_{\tilde{r}}^2$ are diagonal matrices
with their diagonal elements
representing the mass squares of $(\tilde{e}_L,\tilde{\mu}_L,\tilde{\tau}_L)$
and $(\tilde{e}_R,\tilde{\mu}_R,\tilde{\tau}_R)$ respectively.
$Z_L$ and $Z_R$ represent the mixing matrices in the left- and right-handed
sleptons. In this work we consider the mixing between the second
and the third generations (Thereafter we can completely ignore the first
generation). $Z_L$ is then given by
\begin{equation}
Z_L=\left( \begin{array}{cc}  c_L & s_L \\
  -s_L &c_L \end{array}\right)\ ,\ \text{with}\ \ c_L=\cos\theta_L,\ \ s_L=\sin\theta_L\ ,
\end{equation}
while $m_L^2$ is given by
\begin{equation} 
m_L^2 = \left( \begin{array}{cc} m_2^2 & \\ & m_3^2 \end{array}\right)\ .
\end{equation}
On this basis the sneutrino mass matrix can be written as
\begin{equation}
\label{msneu}
M_{\tilde{\nu}}^2=Z_L m_{\tilde{l}}^2 Z_L^\dagger + \frac{1}{2}\cos2\beta M_Z^2\ \ .
\end{equation}

Before giving the relations between $\delta a_\mu$ and Br($\tau\to\mu\gamma$)
in the next section, we first give an analysis of the 
flavor structure of the form factors
$A_L$ and $A_R$ in the following. From Eqs. (\ref{aln})-(\ref{imp}), we
can see that the flavor structure of $A_{L(R)}$ is approximately proportional
to $(M_{\tilde{\nu}}^2)_{ij}^{-1}$ or $(M_{\tilde{l}}^2)_{ij}^{-1}$, 
noticing that the functions $F_i(k_{\alpha a})$
are quite flat in an appropriate range of $k_{\alpha a}$. The numerical
results in the next section justifies our analysis given 
here\footnote{In other works, such as in Ref\cite{pre}, similar relations
are given under the approximation that all the SUSY particles are degenerate.}.
It is easy to get, on the interaction basis,
\begin{equation} 
(M_{\tilde{\nu}}^2)^{-1} =\left( \begin{array}{cc}
\frac{c_L^2}{m_2^2}+\frac{s_L^2}{m_3^2} & c_Ls_L\frac{m_2^2-m_3^2}{m_2^2m_3^2} \\
c_Ls_L\frac{m_2^2-m_3^2}{m_2^2m_3^2} & \frac{s_L^2}{m_2^2}+\frac{c_L^2}{m_3^2}
\end{array}  \right)\ \ .
\end{equation}
$\delta a_\mu$ is approximately proportional to the inverse mass square
of $\tilde{\nu}_\mu-\tilde{\nu}_\mu$,
which is noted as
\begin{equation}
F(\tilde{\nu}_\mu-\tilde{\nu}_\mu)=\frac{c_L^2}{m_2^2}+\frac{s_L^2}{m_3^2}\ , 
\end{equation}
while $\tau\to\mu\gamma$
is approximately proportional to that of $\tilde{\nu}_\tau-\tilde{\nu}_\mu$, 
which is 
\begin{equation}
F(\tilde{\nu}_\tau-\tilde{\nu}_\mu)=
\frac{1}{2}\sin2\theta_L\frac{m_2^2-m_3^2}{m_2^2m_3^2}\ . 
\end{equation}

We consider the following two limit cases:
\begin{equation}
F(\tilde{\nu}_\mu-\tilde{\nu}_\mu)\rightarrow \left\{
\begin{array}{l} \frac{1}{m^2},\ \ \text{if} \ \ m_2^2\approx m_3^2 \approx m^2 \\
\frac{s_L^2}{m_3^2},\ \ \text{if} \ \ m_2^2 \gg m_3^2
\end{array} \right.\ \ ,
\end{equation}
while
\begin{equation}
F(\tilde{\nu}_\tau-\tilde{\nu}_\mu)\rightarrow  \left\{
\begin{array}{l} \frac{1}{2}\sin2\theta_L\frac{\Delta m^2}{m^4},\ \ \text{if} \ \ m_2^2\approx m_3^2 \approx m^2\\
\frac{1}{2}\sin2\theta_L\frac{1}{m_3^2},\ \ \text{if} \ \ m_2^2 \gg m_3^2
\end{array} \right.\ \ .
\end{equation}
In the first case with $m_2^2\approx m_3^2=m^2$, 
we can see that $\delta a_\mu$ 
does not depend on the mixing angle $\theta_L$ and has no direct
relation with Br$(\tau\to\mu\gamma)$. Thus models with gravity or
gauge mediated supersymmetry breaking may predict that $\delta a_\mu$ has
nothing to do with the mixing angle $\theta_L$, as already noticed in Ref\cite{bi}. Thus, to study $\delta a_\mu$ in the first case is 
actually equivalent to the case
of no lepton flavor mixing in the soft sector, which has been extensively 
studied in the literature\cite{susyg-2}. The second case leads us
to the effective SUSY scenario\cite{effe}, 
where the first two generations' sfermions
are as heavy as about $20 TeV$ while the 3rd generation's 
sfermions are kept in a few hundred GeV. 
In this case $\delta a_\mu$ and Br$(\tau\to\mu\gamma)$ 
are closely related; Increase
$\theta_L$ to enhance $\delta a_\mu$ will unavoidably lead to
large  Br$(\tau\to\mu\gamma)$. We have to consider the two quantities simultaneously
and take the experimental bound on  Br$(\tau\to\mu\gamma)$ into account.

The inverse of the mass square of sleptons, on the interaction
basis, is approximately given by
\begin{equation}
(M_{\tilde{l}}^2)^{-1} \approx \left( \begin{array}{cc} A&C\\ C^\dagger&B\end{array}\right)\ ,
\end{equation}
with
\begin{equation}
A\approx (M_{\tilde{\nu}}^2)^{-1},\ \ B\approx A(\theta_L\to\theta_R)
\end{equation}
and
\begin{equation}
C\approx m_\tau\mu\tan\beta\frac{m_2^2-m_3^2}{m_2^2m_3^2}\cdot 
\left[ \begin{array}{ll} \frac{1}{4}\sin2\theta_L\sin2\theta_R
\frac{m_2^2-m_3^2}{m_2^2m_3^2} & \frac{1}{2}\sin2\theta_L
\left( \frac{s_R^2}{m_2^2}+\frac{c_R^2}{m_3^2} \right) \\
\frac{1}{2}\sin2\theta_R \left( \frac{s_L^2}{m_2^2}+\frac{c_L^2}{m_3^2}\right) &
\left(\frac{s_L^2}{m_2^2}+\frac{c_L^2}{m_3^2} \right)
\left( \frac{s_R^2}{m_2^2}+\frac{c_R^2}{m_3^2}\right)
 \end{array} \right]\ \ .
\end{equation}
In matrix $C$ we have omitted the terms proportional to $m_\mu$.
From the above expressions we know 
$F(\tilde{\mu}_L-\tilde{\mu}_L)$ is the same as that of
$F(\tilde{\nu}_\mu-\tilde{\nu}_\mu)$, while
$F(\tilde{\mu}_R-\tilde{\mu}_R)$ is gotten by changing $\theta_L$ to
$\theta_R$ in $F(\tilde{\nu}_\mu-\tilde{\nu}_\mu)$. 
The most interesting result is that of 
$F(\tilde{\mu}_L-\tilde{\mu}_R)$, given by
\begin{equation}
\label{flr}
F(\tilde{\mu}_L-\tilde{\mu}_R)=
\frac{1}{4}m_\tau\mu\tan\beta\sin2\theta_L\sin2\theta_R
\left(\frac{m_2^2-m_3^2}{m_2^2m_3^2}\right)^2\ \ .
\end{equation}
The other two quantities related to $\tau\to\mu\gamma$ are
\begin{equation}
F(\tilde{\mu}_L-\tilde{\tau}_R)=\frac{1}{2} m_\tau\mu\tan\beta\sin2\theta_L
\left(\frac{m_2^2-m_3^2}{m_2^2m_3^2}\right)
\left( \frac{s_R^2}{m_2^2}+\frac{c_R^2}{m_3^2} \right) \ ,
\end{equation}
and
\begin{equation}
F(\tilde{\mu}_R-\tilde{\tau}_L)=\frac{1}{2}m_\tau\mu\tan\beta 
\sin2\theta_R 
\left(\frac{m_2^2-m_3^2}{m_2^2m_3^2}\right)
\left( \frac{s_L^2}{m_2^2}+\frac{c_L^2}{m_3^2}\right)\ ,
\end{equation}
respectively. 

Similarly, we consider the limit case of nearly degenerate sleptons,
$m_2^2\approx m_3^2\approx m^2$. The other term, omitted in
Eq. ({\ref{flr}), which is proportional to $m_\mu$, may become
important. Then we have
\begin{equation}
F(\tilde{\mu}_L-\tilde{\mu}_R)\approx
\frac{\mu\tan\beta}{m^4}\left[ m_\mu+\frac{1}{4}
m_\tau\sin2\theta_L\sin2\theta_R
\left(\frac{\Delta m^2}{m^2}\right)^2\right]
\approx \frac{\mu\tan\beta}{m^4}m_\mu \ .
\end{equation}
We can thus reach the same conclusion as before, that is,
if $m_2^2\approx m_3^2\approx m^2$, 
$\delta a_\mu$ has no direct relation to slepton mixing angles.
In case of $m_2\gg m_3$, we have
\begin{equation}
F(\tilde{\mu}_L-\tilde{\mu}_R)\approx
\frac{\mu\tan\beta}{m_3^4}\left[ m_\mu s_L^2s_R^2 +
\frac{1}{4}m_\tau \sin2\theta_L\sin\theta_R\right]\ ,
\end{equation}
where the second term may dominate.
In this case $\delta a_\mu$ depends crucially on
 the mixing angles $\theta_L$ and $\theta_R$.

The key feature of the effective Lagrangian in Eq. (\ref{eff}) is
that there is a chiral flip between the initial and final fermion
states. This feature leads to that, in the case of no slepton mixing,
all terms in Eqs. (\ref{auneu})
and (\ref{aucha}) will produce at least one muon mass, 
$m_\mu$, suppression, which either
comes from the mass insertion on the external fermion legs, 
or from the Yukawa coupling
vertices, or from the left- and right-handed smuon mixing. This
can be explicitly examined by checking all the Feynman diagrams
on the interaction basis.
The quite interesting point in the case of slepton mixing is that 
$F(\tilde{\mu}_L-\tilde{\mu}_R)$, given in Eq. (\ref{flr}),
is approximately proportional to
$m_\tau$, which can give an enhancement to $\delta a_\mu$. 
This term may dominate
others if both the left- and right-handed mixing
angles are large. We will show this point in the next
section.

From the above analysis we have shown that by changing 
to the interaction basis, 
$\delta a_\mu$ and Br$(\tau\to\mu\gamma)$ 
can manifest their  dependence on 
the SUSY and mixing parameters.
This basis becomes very convenient for our discussion of
the relation between $\delta a_\mu$ and Br$(\tau\to\mu\gamma)$
later.

It should be mentioned that the parameters $m_{2,3}$ are different
for sleptons and sneutrinos, as shown in Eqs. (\ref{mslp}-\ref{mslpr}) 
and (\ref{msneu}). We adopt the same symbol in
$F(\tilde{\nu}-\tilde{\nu})$ and $F(\tilde{\mu}-\tilde{\mu})$
only for simplicity. In numerical
calculations we adopt the full form in Eqs. (\ref{mslp}) and (\ref{msneu}).

\section{ Bound on $\delta a_\mu$ and numerical results }

In this section, we focus our discussion on the effective
SUSY scenario, i.e., $m_2\gg m_3$. In this case, $\delta a_\mu$
and Br$(\tau\to\mu\gamma)$ are closely related, as shown
in the last section. We will give an approximate bounds 
on $\delta a_\mu$ through analytic relations between
$\delta a_\mu$ and Br$(\tau\to\mu\gamma)$, taking into 
account the experimental up limit on the LFV processes.
Numerical results are also presented.

The free parameters in this calculation are the
Higgsino mass, $\mu$, U(1)$_Y$ and SU(2)$_W$ gaugino masses, 
$M_1$ and $M_2$,
ratio of VEVs, $\tan\beta$, mixing angles, $\theta_L$ and $\theta_R$,
slepton mass squares, $m^2_{\tilde{l}_2}$, $m^2_{\tilde{l}_3}$,
$m^2_{\tilde{r}_2}$, $m^2_{\tilde{r}_3}$, and the trilinear terms $A_l$.
Throughout the whole calculation we fix $A_l=0$,  
$m^2_{\tilde{l}_2}=m^2_{\tilde{r}_2}=20TeV$. If we do not state
explicitly we will take the relation $M_1=\frac{5\alpha_1}{3\alpha_2}M_2$
and $m^2_{\tilde{l}_3}=m^2_{\tilde{r}_3}$.
We demand all the SUSY particle spectra be above the present experimental
lower limit. 

\subsection{$\delta a_\mu$ with $\theta_R=0$ }

\begin{center}
\begin{figure}
\begin{picture}(500,250)(20,50)
\ArrowLine(145,200)(190,200)
\ArrowLine(190,200)(310,200)
\ArrowLine(310,200)(355,200)
\DashArrowArcn(250,200)(60,180,0){4}
\Photon(265,170)(335,120){4}{7}
\Text(200,250)[r]{\Large $\tilde{\nu}_\mu(\tilde{\nu}_\tau)$}
\Text(305,250)[l]{\Large $\tilde{\nu}_\mu$}
\Text(155,185)[]{\Large $\mu_R(\tau_R)$}
\Text(345,185)[]{\Large $\mu_L$}
\Text(220,185)[r]{\Large $\widetilde{H}_D$}
\Text(250,185)[r]{\Large $\widetilde{H}_U$}
\Text(260,185)[l]{\Large $\widetilde{W}_L$}         
\Text(290,185)[l]{\Large $\widetilde{W}_R$}
\Text(220,212)[]{\Large $\mu$}
\Text(280,212)[]{\Large $M_2$}
\Vertex(220,200){3}
\Vertex(250,200){3}
\Vertex(280,200){3}
\end{picture}

\vspace*{-1.5cm}
\caption{\label{fig2} Feynman diagram which gives the dominant
contribution to $\delta a_\mu$ (and to the process $\tau \to\mu\gamma$)
in case of only left-handed slepton mixing. The black dots in the chargino
line are mass insertions, with the middle dot representing $\sqrt{2}M_W\sin\beta$.}
\end{figure}
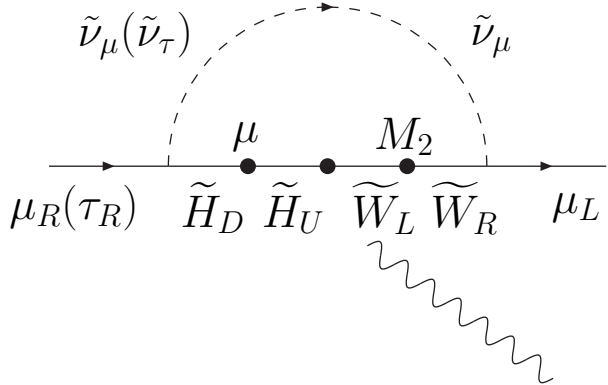
\end{center}

When there is only left-handed mixing in the slepton sector, 
the most important
contribution to $\delta a_\mu$ and Br$(\tau\to\mu\gamma)$ 
comes from the diagram in FIG. \ref{fig2}, given on the
interaction basis. From this diagram
we can directly read that
\begin{equation}
A_R^{23}(c)\approx\frac{1}{2}\delta a_\mu^{(c)}/m_\mu^2
\frac{F(\tilde{\nu}_\tau-\tilde{\nu}_\mu)}{F(\tilde{\nu}_\mu-\tilde{\nu}_\mu)}
\approx \frac{1}{2}\delta a_\mu^{(c)}/m_\mu^2\frac{c_L}{s_L}\ \ .
\end{equation}
Then we have, assuming $\theta_L=\pi/4$, that
\begin{eqnarray}
Br(\tau\to\mu\gamma)&\approx & \frac{\alpha_{em}}{4}m_\tau^5|A_R^{23}(c)|^2/\Gamma_\tau \nonumber \\
&\approx & \frac{\alpha_{em}}{4}m_\tau^5/\Gamma_\tau\left| 
\frac{\delta a_\mu^{(c)}}{2m_\mu^2}\frac{c_L}{s_L}
\right|^2\nonumber \\
&\approx & 2.9\times 10^{13} |\delta a_\mu|^2\ .
\end{eqnarray}
From the present experimental upper bound on 
Br$(\tau\to\mu\gamma)$( $< 10^{-6}$\cite{bound}),
we get that 
\begin{equation}
\delta a_\mu < 1.9 \times 10^{-10},\ \ \text {in case of}\ \ \theta_R=0\ .
\end{equation}

From this diagram we also have
the conclusion that
\begin{equation}
\mu M_2 > 0, \ \ \text {in case of}\ \ \theta_R=0\ 
\end{equation}
in order that SUSY gives positive contribution to  $\delta a_\mu$.
The same diagram gives the dominant contribution to $\delta a_\mu$
in the case of no lepton flavor mixing. Thus the same conclusion
of the sign of $\mu$ is given in that case.

Numerical study verifies our above estimation.

\subsection{$\delta a_\mu$ with $\theta_L=0$ } 

\begin{center}
\begin{figure}
\begin{picture}(500,250)(20,50)
\ArrowLine(145,200)(190,200)
\ArrowLine(190,200)(310,200)
\ArrowLine(310,200)(355,200)
\DashArrowArcn(250,200)(60,180,0){4}
\Photon(280,270)(350,305){4}{7}
\Text(200,250)[r]{\Large $\tilde{\mu}_R(\tilde{\tau}_R)$}
\Text(305,250)[l]{\Large $\tilde{\mu}_R$}
\Text(155,185)[]{\Large $\mu_R(\tau_R)$}
\Text(345,185)[]{\Large $\mu_L$}
\Text(220,185)[r]{\Large $\widetilde{B}_L$}
\Text(250,185)[r]{\Large $\widetilde{B}_R$}
\Text(260,185)[l]{\Large $\widetilde{H}_U$}
\Text(290,185)[l]{\Large $\widetilde{H}_D$}
\Text(220,212)[]{\Large $M_1$}
\Text(280,212)[]{\Large $-\mu$}
\Vertex(220,200){3}
\Vertex(250,200){3}
\Vertex(280,200){3}
\end{picture}

\vspace*{-3.cm}
\caption{\label{fig3} Feynman diagram which gives the dominant
contribution to $\delta a_\mu$ (and to the process $\tau \to\mu\gamma$)
in case of only right-handed slepton mixing. The black dots in the neutralino
line are mass insertions, with the middle dot representing $M_Z\sin\beta\sin\theta_W$.}
\end{figure}
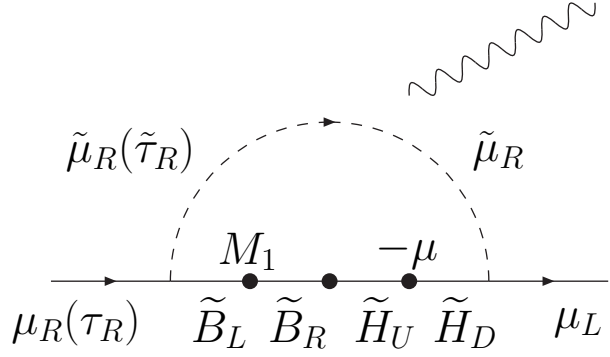
\end{center}

In case of only right-handed mixing, the chargino-sneutrino diagram
gives no contribution to $\delta a_\mu$. The most important
contribution to $\delta a_\mu$ and Br$(\tau\to\mu\gamma)$
comes from the diagram in FIG. \ref{fig3}, given on the
interaction basis. 
If we ignore the mixing between the left- and right-handed sleptons,
$Z_R$ is approximately the slepton mixing matrix.
From FIG. \ref{fig3} we then have 
\begin{equation}
\label{rmix}
A_R^{23}(n)\approx \frac{1}{2}\frac{\delta a_\mu^{(n)}}{m_\mu^2}
\left(\frac{m_\mu}{m_\tau}\right)
\frac{F(\tilde{\tau}_R-\tilde{\mu}_R)}{F(\tilde{\mu}_R-\tilde{\mu}_R)} 
\approx \frac{1}{2}\frac{\delta a_\mu^{(n)}}{m_\mu^2}
\left(\frac{m_\mu}{m_\tau}\right)\frac{c_R}{s_R}\ \ .
\end{equation}

Then we have, assuming $\theta_R=\pi/4$, that
\begin{eqnarray}
Br(\tau\to\mu\gamma)&\approx & \frac{\alpha_{em}}{4}m_\tau^5|A_R^{23}(n)|^2/\Gamma_\tau \nonumber \\
&\approx & \frac{\alpha_{em}}{4}m_\tau^5/\Gamma_\tau\left|
\frac{\delta a_\mu^{(n)}}{2m_\mu^2}\frac{m_\mu}{m_\tau}\frac{c_R}{s_R}
\right|^2\nonumber \\
&\approx & 1.\times 10^{11} |\delta a_\mu|^2\ .
\end{eqnarray}
From the present upper limit of $Br(\tau\to\mu\gamma) < 10^{-6}$,
we get that
\begin{equation}
\label{bdr}
\delta a_\mu < 32 \times 10^{-10},\ \ \text {in case of}\ \ \theta_L=0\ .
\end{equation}

This upper bound is much larger than that in the case of only left-handed
mixing. It is obvious that the factor $\frac{m_\mu}{m_\tau}$ in 
Eq. (\ref{rmix}), which greatly suppresses Br($\tau\to\mu\gamma$), 
helps to increase the bound.
This factor comes  from the $\mu_L-\widetilde{H}_D-\tilde{\mu}_R$
Yukawa coupling vertex in FIG. \ref{fig3}, where the Higgsino 
component $\widetilde{H}_D$ has to be associated
with the muon line since there is only right-handed 
mixing in the slepton sector.
However, in  FIG. \ref{fig2}, where the charged
Higgsino component $\widetilde{H}_D$ is associated with
the tau line, no such factor helps to suppress Br($\tau\to\mu\gamma$).

Another interesting point is that the mass insertion for the
neutral component of  $\widetilde{H}_U \widetilde{H}_D$ is
$-\mu$, while it is $\mu$ for the same term of the charged component.
These terms are clearly shown in the mass matrices of charginos and neutralinos
in the Appendix. 
This sign difference comes in when we contract the SUSY invariant
term $\mu\epsilon_{ab}H_D^aH_U^b$. Thus we have 
\begin{equation}
\mu M_1 < 0, \ \ \text {in case of}\ \ \theta_L=0\
\end{equation}
to give positive contribution to  $\delta a_\mu$.
This means that if we set $M_1$ and $M_2$ have same sign,
which is well motivated theoretically, $\mu$ should be
negative in this case. We have numerically demonstrated this point
by changing the signs of $M_1$ and $\mu$ simultaneously 
and finding that $\delta a_\mu$ almost has the same value.

\begin{figure}
\includegraphics[scale=0.6]{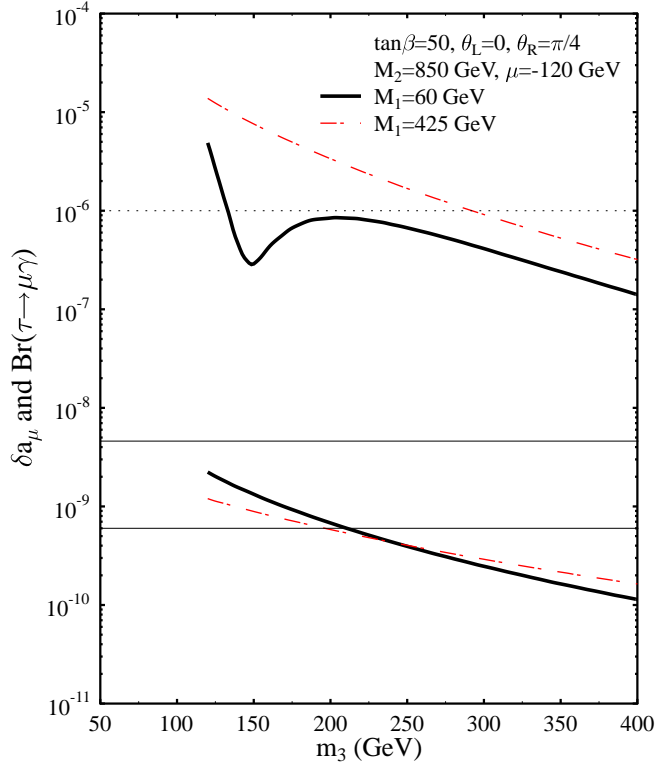}
\caption{\label{fig4} 
$\delta a_\mu$ and Br($\tau\to\mu\gamma$) as functions of 
$m_3=m_{\tilde{l}_3}=m_{\tilde{r}_3}$ in
case of $\theta_L=0$, $\theta_R=\pi/4$. The other parameters are
$\tan\beta=50$, $M_2=850 GeV$. $M_1$ is fixed to be $60 GeV$ for
the solid line and $0.5 M_2$ for the dashed line.
The horizontal lines represent the E821 $\pm 2\sigma$ 
bounds (solid) and the upper limit of Br($\tau\to\mu\gamma$) (dotted). }
\end{figure}

Since we ignored the left-right mixing  between the sleptons,
the naive bound we get in Eq. (\ref{bdr}) should be examined numerically.
The numerical results in this case are shown in FIG. \ref{fig4}.
In this figure (and all similar figures below) we
draw $\delta a_\mu$ and Br($\tau\to\mu\gamma$) in the same
figure as functions of 
$m_{\tilde{l}_3}=m_{\tilde{r}_3}=m_3$. 
The upper group of curves represent Br($\tau\to\mu\gamma$)
while the corresponding curve in the lower group is
$\delta a_\mu$ with same parameters.
The two solid horizontal lines represent the E821 $\pm 2\sigma$ bounds,
$\delta a_\mu = 6,\ 46\times 10^{-10}$. The dotted horizontal line
is the experimental upper bound on branching ratio of $\tau\to\mu\gamma$, 
Br($\tau\to\mu\gamma$)=$10^{-6}$.

We take large $\tan\beta$(=$50$) and $M_2$(=$850 GeV$),
while $\mu$ is negative.
If we adopt the GUT motivated relation 
$M_1=\frac{5\alpha_1}{3\alpha_2}M_2\approx 0.5M_2$
we have $\delta a_\mu < 3\times 10^{-10}$ to satisfy the 
Br($\tau\to\mu\gamma$) bound.
However, if we relax the above relation and fix $M_1=60 GeV$, $\delta a_\mu$
can be as large as $\sim 17\times 10^{-10}$ without violating
the bound of Br($\tau\to\mu\gamma$). This case corresponds to
that the LSP (lightest supersymmetric particle) is bino, which
is much lighter than other neutralinos.

\subsection{$\delta a_\mu$ with no $\theta=0$ }

\begin{center}
\begin{figure}
\begin{picture}(500,250)(20,50)
\ArrowLine(145,200)(190,200)
\ArrowLine(190,200)(310,200)
\ArrowLine(310,200)(355,200)
\DashArrowArcn(250,200)(60,180,0){4}
\Photon(290,270)(360,305){4}{7}
\Text(155,185)[]{\Large $\mu_R$}
\Text(345,185)[]{\Large $\mu_L$}
\Text(220,185)[]{\Large $\widetilde{B}_L$}
\Text(290,185)[]{\Large $\widetilde{B}_R$}
\Text(180,220)[]{\Large $\tilde{\mu}_R$}
\Text(310,217)[l]{\Large $\tilde{\mu}_L$}
\Text(200,255)[]{\Large $\tilde{\tau}_R$}
\Text(292,252)[l]{\Large $\tilde{\tau}_L$}
\Text(250,273)[]{\large $m_\tau \mu\tan\beta$}
\Vertex(250,200){3}

\Vertex(250,260){3}
\Vertex(198,230){3}
\Vertex(302,230){3}

\end{picture}

\vspace*{-3.cm}
\caption{\label{fig5} 
Feynman diagram which gives the dominant
contribution to $\delta a_\mu$ 
in the case that both the left- and right-handed slepton mixing
are large. }
\end{figure}
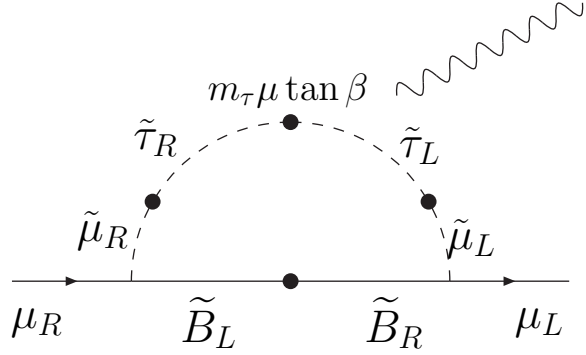
\end{center}

This case is the most general and most interesting one. 
Our numerical calculation 
mainly focus on this case.
In this case we have derived in Eq. (\ref{flr})
that there is an $m_\tau$ enhancement
to $F(\tilde{\mu}_L-\tilde{\mu}_R)$
if both the left- and right-handed mixing is large in the
slepton sector. The enhancement leads to that the diagram in FIG. \ref{fig5}
may give dominant contribution to $\delta a_\mu$ if both $\theta_L$ and
 $\theta_R$ are large. However, there is no obvious
term which give dominant contribution to Br($\tau\to\mu\gamma$).
We find that in small $m_3$ region the diagram in FIG. \ref{fig5} with $\mu_R$ 
replaced by $\tau_R$ may
dominates other terms to Br($\tau\to\mu\gamma$). 
In this case we get a similar
limit as that given in the case with only right-handed mixing, i.e.,
\begin{equation}
\delta a_\mu < 32\times 10^{-10} \ \ \ \text{in case of no}\ \ \theta=0\ \ .
\end{equation}
However, this bound is very loose because in large parameter space
the contribution to Br($\tau\to\mu\gamma$) by exchanging $\chi^\pm$
is more important than that by exchanging $\chi^0$. We can not
get simple relation between $\delta a_\mu$ and 
Br($\tau\to\mu\gamma$) for this case and have to
study it numerically. 

\begin{figure}
\includegraphics[scale=0.6]{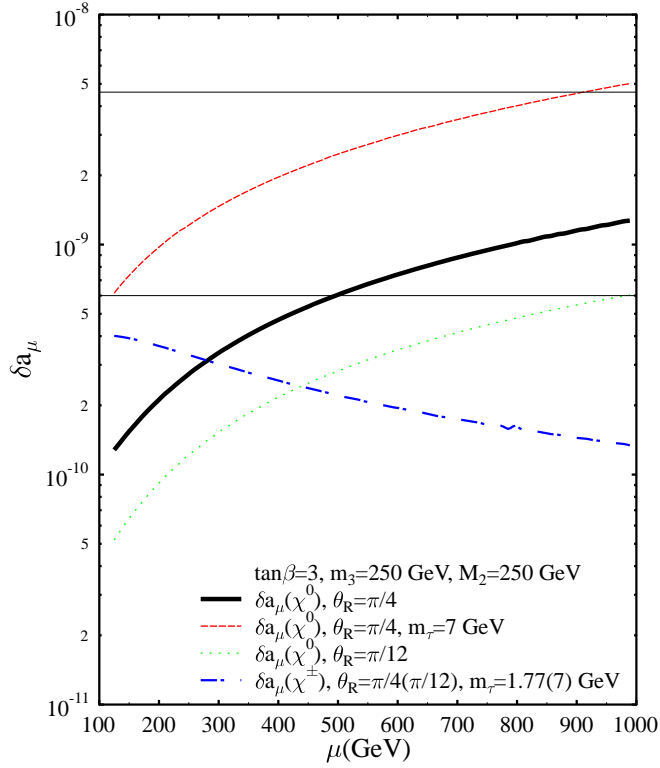}
\caption{\label{fig6}
$\delta a_\mu$ as function of $\mu$ for $\tan\beta=3$, 
$M_2=250 GeV$, $m_3=250 GeV$ and $\theta_L=\pi/4$. 
$\delta a_\mu(\chi^0)$ and $\delta a_\mu(\chi^\pm)$ represent
the contribution coming from exchanging neutralino and
chargino respectively.
The horizontal lines represent the E821 $\pm 2\sigma$
bounds of $\delta a_\mu$. }
\end{figure}

At first we will numerically verify that FIG. \ref{fig5} indeed gives
important contribution to $\delta a_\mu$ by
displaying the contributions from exchanging $\chi^0$ and
$\chi^\pm$ separately in FIG. \ref{fig6}. 
In order to show the $m_\tau$ enhancement
and the dependence on the mixing angles, 
we plot another two lines
for setting $m_\tau=7 GeV$ in the slepton
mass matrix and for $\theta_R=\pi/12$. 
We notice that $\delta a_\mu$ changes linearly as
$m_\tau$, demonstrating the term proportional to $m_\tau$
indeed gives dominant contribution to $\delta a_\mu$.
However, contribution from exchanging charginos has no change
by changing values of $m_\tau$ and $\theta_R$.
From Eq. (\ref{flr}) it is obvious that the neutralino contribution
becomes large as $\mu$ increases, while the chargino contribution
becomes small since large $\mu$ leads to heavy chargino mass.

\begin{figure}
\includegraphics[scale=0.6]{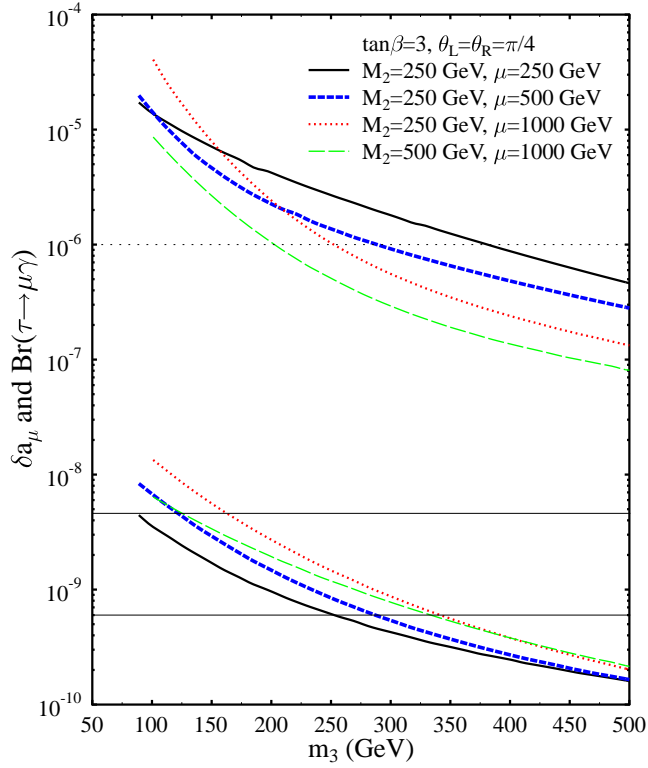}
\caption{\label{fig7}
$\delta a_\mu$ and Br($\tau\to\mu\gamma$) as functions of $m_3$ for
$\theta_L=\pi/4$, $\theta_R=\pi/4$,
$\tan\beta=3$, $M_2=250, 500 GeV$ and $\mu=250, 500, 1000 GeV$.
The horizontal lines represent the E821 $\pm 2\sigma$
bounds of $\delta a_\mu$ (solid) and the upper 
limit of Br($\tau\to\mu\gamma$) (dotted). }
\end{figure}

FIG. \ref{fig7} displays $\delta a_\mu$ and Br($\tau\to\mu\gamma$)
for $\tan\beta=3$, $\theta_L=\theta_R=\pi/4$. If both $M_2$ and $\mu$ are large,
there is a large region which can accommodate 
$\delta a_\mu$ and Br($\tau\to\mu\gamma$) simultaneously.
As $\mu$ becomes large,  Br($\tau\to\mu\gamma$) decreases
while $\delta a_\mu$ increases. This is understood that
large $\mu$ enhances $F(\tilde{\mu}_L-\tilde{\mu}_R)$ 
and leads to large chargino mass, which decreases Br($\tau\to\mu\gamma$).

\begin{figure}
\includegraphics[scale=0.6]{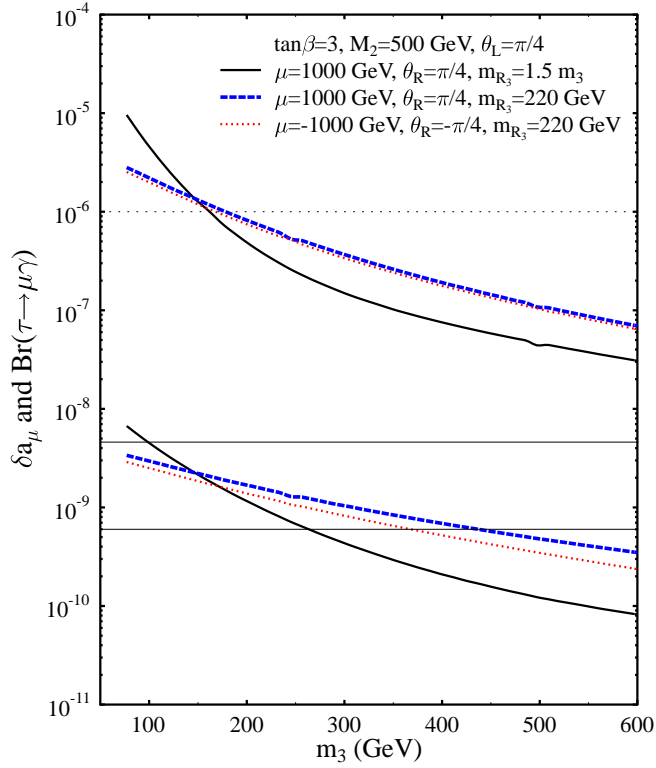}
\caption{\label{fig8}
$\delta a_\mu$ and Br($\tau\to\mu\gamma$) as functions of 
$m_{\tilde{l}_3}=m_3$ for
$\tan\beta=3$, $M_2=500 GeV$ and $\theta_L=\pi/4$. 
Taking $m_{\tilde{r}_3}=1.5 m_3,\ 220 GeV$, $\mu=\pm 1000 GeV$
and $\theta_R=\pm\pi/4$ respectively.
The horizontal lines represent the E821 $\pm 2\sigma$
bounds of $\delta a_\mu$ (solid) and the upper
limit of Br($\tau\to\mu\gamma$) (dotted). }
\end{figure}

In FIG. \ref{fig8} we relax the relation 
$m_{\tilde{l}_3}=m_{\tilde{r}_3}$.
We show $\delta a_\mu$ and Br($\tau\to\mu\gamma$) 
as functions of $m_{\tilde{l}_3}=m_3$ 
for $m_{\tilde{r}_3}=1.5 m_3$
and $m_{\tilde{r}_3}=220 GeV$.
From Eq. (\ref{flr}) we notice that the sign
of $\mu$ can be either positive or negative depending
on the relative sign of $\theta_L$ and $\theta_R$.
We also plot $\delta a_\mu$ and Br($\tau\to\mu\gamma$)
for changing the sign of $\mu$ and $\theta_R$ simultaneously in
 FIG. \ref{fig8}. There is little effect on $\delta a_\mu$ by
the sign reverse.

\begin{figure}
\includegraphics[scale=0.6]{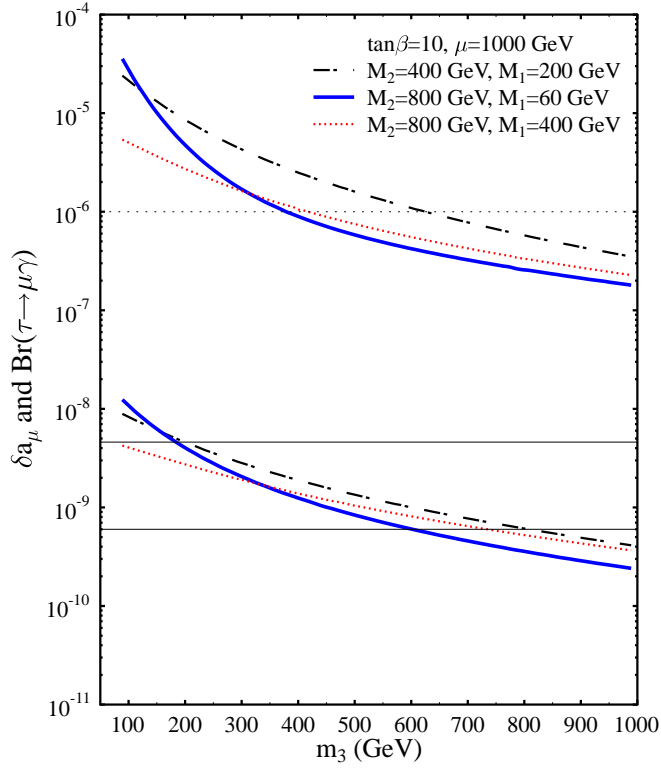}
\caption{\label{fig9}
$\delta a_\mu$ and Br($\tau\to\mu\gamma$) as functions of 
$m_{\tilde{l}_3}=m_3$ for
$\tan\beta=10$, $\mu=1000 GeV$, $\theta_L=\theta_R=\pi/4$,
 $m_{\tilde{r}_3}=300 GeV$, $M_2=400, 800 GeV$ and $M_1=0.5 M_2, 60 GeV$.
The horizontal lines represent the E821 $\pm 2\sigma$
bounds of $\delta a_\mu$ (solid) and the upper
limit of Br($\tau\to\mu\gamma$) (dotted). }
\end{figure}

Since Br($\tau\to\mu\gamma$) is approximately proportional to
$\tan^2\beta$, its upper limit constrains $\tan\beta$ strongly.
In above figures we take $\tan\beta=3$. In FIG. \ref{fig9}
we plot $\delta a_\mu$ and Br($\tau\to\mu\gamma$)
 as functions of $m_{\tilde{l}_3}=m_3$ 
for $\tan\beta=10$, and $\mu=1000 GeV$, $M_2=400,\ 800 GeV$,
$M_1=\frac{5\alpha_1}{3\alpha_2}\approx 0.5 M_2,\ 60 GeV$ respectively. 
We fix $m_{\tilde{r}_3}=300 GeV$ in this figure.
We can see that when $\tan\beta$ is as large as 10
there is still a large region for $m_3$ to accommodate
$\delta a_\mu$ and  Br($\tau\to\mu\gamma$) simultaneously
if $M_2$ is large.
Generally $m_3$ takes larger value than that in case of
$\tan\beta=3$ to satisfy the $\delta a_\mu$ and  Br($\tau\to\mu\gamma$) 
bounds. We also notice that relaxing the
relation $M_1=0.5 M_2$ does not change the result much.

In summary, in the effective SUSY scenario,
when both the left- and right-handed slepton
mixing is large, SUSY can enhance $\delta a_\mu$ to within
the E821 $\pm 2\sigma$ bounds in a large parameter space,
which can satisfy the constraints by experimental limit on
Br($\tau\to\mu\gamma$).
In this case small $\tan\beta$ is slightly favored.
Higgsino mass parameter $\mu$ can be either positive or negative
depending on the relative sign between $\theta_L$ and $\theta_R$.
We find that $\delta a_\mu$ can reach up to $\sim 20\times 10^{-10}$ even
keeping the relation $M_1\approx 0.5 M_2$, implying that bino
is not necessarily kept very light in this case.

\section{Summary and conclusions}

In this work we study the correlation between the SUSY
contribution to $\delta a_\mu$ and Br($\tau\to\mu\gamma$)
by translating the analytic expressions in the
mass eigenstates to interaction basis, where the
gauge coupling and Yukawa interaction vertices are
all diagonal.  If the slepton mass eigenstates are
approximately degenerate, $\delta a_\mu$ does not depend on the 
lepton flavor mixing angles and has no direct relation
with the LFV processes. In this case, the analysis of
$\delta a_\mu$ is actually the same 
as no slepton mixing, as most authors done in the literature.
Another case that the second (and first) generation
slepton is much heavier than the third generation
slepton corresponds to the scenario of effective SUSY.
We mainly investigate this case in our work.

In the effective SUSY scenario,
if there is only left-handed mixing on the slepton sector,
the upper limit of Br($\tau\to\mu\gamma$) constrains
SUSY contribution to $a_\mu$ smaller than $1.9\times 10^{-10}$.
In the case of only right-handed mixing on the slepton sector,
numerical study shows that $\delta a_\mu$ can be at most
$\sim 17\times 10^{-10}$ if bino is the LSP and much
lighter than other neutralinos. In this case the Higgsino mass 
$\mu$ is negative relative to $M_1$ in order to give positive
contribution to $a_\mu$. In the case of both left- and
right-handed mixing angles being large, we find the
diagram exchanging bino can give dominant
contribution to  $\delta a_\mu$. The sign of $\mu$
is determined by making this diagram positive.
Thus $\mu$ can be either positive or negative
depending on the relative sign between the left- and
right-handed slepton mixing angles. Numerical study
shows that in this case there is large parameter space
accommodating $\delta a_\mu$ and Br($\tau\to\mu\gamma$) simultaneously.
The SUSY contribution to  $a_\mu$
can reach up to $\sim 20\times 10^{-10}$, without
requiring a very light bino.

Our study shows that the parameter space is quite different
in the effective SUSY scenario compared with 
that in the case of no slepton mixing. The small $\tan\beta$
value is more favored. The sign of $\mu$ is not constrained
by the ($g_\mu-2$) experiment. Finally the effective SUSY 
scenario can not be excluded by the E821 experiment
if we take the lepton flavor mixing effects into account. 

\appendix*
\section{}

In this appendix we present our conventions for the
SUSY parameters and some analytic expressions for $\delta a_\mu$
and Br($\tau\to\mu\gamma$). 
For most part we adopt the conventions
given in Ref. \cite{rosiek}. 

The 2-component charged Higgsinos, 
$\widetilde{H}_D^2$, $\widetilde{H}_U^1$, and charged winos, 
$\lambda^{\pm}=\frac{1}{\sqrt{2}}(\lambda^{1}_{W}\mp i\lambda^{2}_{W})$, 
combine to give two 4-component Dirac fermions named
charginos, where $\widetilde{H}_D^2$ and $\widetilde{H}_U^1$ are the second and
the first components of the down and up Higgsino SU(2) doublets respectively, 
$\lambda^{1}_{W}$ and $\lambda^{2}_{W}$ are the first and the second
components of the wino SU(2) triplet.
The mass matrix of charginos, given on the interaction eigenstates, is
\begin{equation}
M_{\chi} = \left[\matrix{
{m}_2 & \sqrt{2} M_W \sin\beta \cr
\sqrt{2} M_W \cos\beta & \mu } \right] \ \ .
\end{equation}
The unitary mixing matrices $Z^-$, $Z^+$ satisfy
\begin{equation}
(Z^-)^TM_{\chi} Z^+
= \text {diag}\left(m_{{\chi}_1}, m_{{\chi}_2}\right) \ ,
\end{equation}
which are defined by
\begin{equation}
\left( \begin{array}{c}
 -i\ \lambda^{-}\\ \widetilde{H}_D^2
\end{array} \right)\ = \
Z^{-}\ \left(\begin{array}{c}\varphi^{-}_{1}\\ \varphi^{-}_{2}\end{array}
\right)\ \ ,
\end{equation}
and
\begin{equation}
\left( \begin{array}{c}
 -i\ \lambda^{+}\\ \widetilde{H}_U^1
\end{array} \right)\ = \
Z^{+}\ \left(\begin{array}{c}\varphi^{+}_{1}\\ \varphi^{+}_{2}\end{array}
\right)\ \ .
\end{equation}
The four-component Dirac charginos are defined by $\chi_{i}^{+}=\left[
\begin{array}{c}\varphi^{+}_{i}\\ \bar{\varphi^{-}_{i}}\end{array}\right]$.
The mass term
which will appear in the final form of Lagrangian is $-m_{\chi_i}\bar{\chi_i}
\chi_i$.

The third component of wino, $\lambda_W^3$, bino, $\tilde{B}$, 
and neutral Higgsinos, $\widetilde{H}_D^1$, $\widetilde{H}_U^2$, 
combine to give four Majorana neutralinos.
The mass matrix for neutralinos on the interaction basis is
\begin{equation}
M_{{\chi}^0} = \left[\matrix{
{m}_1   & 0 & -M_Z\cos\beta\sin\theta_W & M_Z\sin\beta\sin\theta_W \cr
0 & {m}_2 & M_Z\cos\beta\cos\theta_W & -M_Z\sin\beta\cos\theta_W \cr
-M_Z\cos\beta\sin\theta_W & M_Z\cos\beta\cos\theta_W & 0 & -\mu\cr
M_Z\sin\beta\sin\theta_W & -M_Z\sin\beta\cos\theta_W & -\mu & 0 } \right] \ , \ \
\end{equation}
which is diagonalized by
\begin{equation}
Z_N^T M_{{\chi}^0} Z_N
= \text {diag}\left(m_{{\chi}^0_1}, m_{{\chi}^0_2},
m_{{\chi}^0_3}, m_{{\chi}_4^0}\right) \ \ .
\end{equation}
The unitary mixing matrix $Z_N$ is defined by
\begin{equation}
\left( \begin{array}{c}
-i\lambda_{B}\\ -i\lambda_{W}^3\\  \widetilde{H}_D^1 \\  
\widetilde{H}_U^2 \end{array}
\right) \ =\ Z_{N}\left( \begin{array}{c}
\varphi^{0}_{1}\\ \varphi^{0}_{2}\\ \varphi^{0}_{3}\\ \varphi^{0}_{4}
\end{array}\right) \ \ .
\end{equation}
The four-component Majorana  neutralinos are given by $\chi^{0}_{i}=
\left[\begin{array}{c}
\varphi^{0}_{i}\\ \bar{\varphi^{0}_{i}}\end{array}\right]$.
The mass term in the final form of Lagrangian of neutralino is $-\frac{1}{2}
m_{\chi^0_i}\bar{\chi^0_i}\chi^0_i$.

The mass matrices for sleptons and sneutrinos have been given in 
Eqs. (\ref{mslp}) and (\ref{msneu}). They are diagonalized by
\begin{equation}
Z_{\tilde{L}}^\dagger M_{\tilde{l}}^2 Z_{\tilde{L}} = \text {diag}(m_{\tilde{l}_\alpha}^2),
\ \alpha=1\cdots 6,\ \ 
\end{equation}
and
\begin{equation}
Z_{\tilde{\nu}}^\dagger M_{\tilde{\nu}}^2 Z_{\tilde{\nu}} = \text{diag}(m_{\tilde{\nu}_\alpha}^2),\ \ \alpha=1,2,3\ .
\end{equation}
The relation between the gauge eigenstates and the mass eigenstates are 
(omitting the first generation)
\begin{equation}
\left( \begin{array}{c} \tilde{\mu}_L \\ \tilde{\tau}_L \\
\tilde{\mu}_R^* \\ \tilde{\tau}_R^* \end{array} \right)
=
Z_{\widetilde{L}} \left( \begin{array}{c} 
m_{\tilde{l}_1}^2 \\ m_{\tilde{l}_2}^2 \\
m_{\tilde{l}_3}^2 \\ m_{\tilde{l}_4}^2
\end{array}\right)
\end{equation}
and similar expression for sneutrinos.

The functions $F_i(k)$ in Eqs. (\ref{auneu})-(\ref{aucha})
and (\ref{aln})-(\ref{imp}) are given as follows.
For neutralino-exchange we have
\begin{eqnarray}
F_1(k)&=&\frac{1-6k+3k^2+2k^3-6k^2\log k}{6(1-k)^4}\ ,\\
F_2(k)&=&\frac{1-k^2+2k\log k}{(1-k)^3}\ \ ,
\end{eqnarray}
with $k_{\alpha a}={m_{\chi_a^0}^2}/{m_{\tilde{l}_\alpha}^2}$. 
For chargino-exchange we have
\begin{eqnarray}
F_3(k)&=&\frac{2+3k-6k^2+k^3+6k\log k}{6(1-k)^4} \ ,\\
F_4(k)&=&\frac{3-4k+k^2+2\log k}{(1-k)^3}\ \ ,
\end{eqnarray}
with $k_{\alpha a}={m_{\chi_a^-}^2}/{m_{\tilde{\nu}_\alpha}^2}$.

\begin{acknowledgments}
This work is supported by the National Natural Science Foundation
of China under the grant No. 10105004.
\end{acknowledgments}

\end{document}